\documentclass[superscriptaddress,aps,prl,twocolumn]{revtex4-1}
\usepackage{graphicx}
\usepackage{textcomp} 
\usepackage[utf8]{inputenc}
\usepackage{amsmath}
\usepackage{amsfonts}
\usepackage{amssymb}
\usepackage{mathbbol}
\usepackage{isomath}
\usepackage{hyperref}
\usepackage{url}

\begin{document}
\title{Cellular velocity, electrical persistence and sensing in developed and vegetative cells during electrotaxis}
\author{Isabella Guido} \email{isabella.guido@ds.mpg.de}  \affiliation{Max Planck Institute for Dynamics and Self-Organization (MPIDS), 37077
	G\"ottingen, Germany}
\author{Douglas Diehl} \affiliation{Max Planck Institute for Dynamics and
	Self-Organization (MPIDS), 37077 G\"ottingen,
	Germany}
\author{Nora Aleida Olszok} \affiliation{Max Planck Institute for Dynamics and Self-Organization (MPIDS), 37077 G\"ottingen, Germany}
\author{Eberhard Bodenschatz} \affiliation{Max Planck Institute for
  Dynamics and Self-Organization (MPIDS), 37077 G\"ottingen,
  Germany}
\affiliation{Institute for Dynamics of Complex Systems,
  Georg-August-University G{\"o}ttingen, 37073 G{\"o}ttingen, Germany}
\affiliation{ Laboratory of Atomic and Solid-State Physics,
  Cornell University, Ithaca, NY 14853, United States} 

\begin{abstract}
\noindent Cells have the ability to detect electric fields and respond to them with directed migratory movement. Investigations identified genes and proteins that play important roles in defining the migration efficiency. Nevertheless, the sensing and transduction mechanisms underlying directed cell migration are still under discussion. We use \textit{Dictyostelium discoideum} cells as model system for studying eukaryotic cell migration in DC electric fields. We have defined the temporal electric persistence to characterize  the memory that cells have in a varying electric field. In addition to imposing a directional bias, we observed that the electric field influences the cellular kinematics by accelerating the movement of cells along their paths. Moreover, the study of vegetative and briefly starved cells provided insight into the electrical sensing of cells. We found evidence that conditioned medium of starved cells was able to trigger the electrical sensing of vegetative cells that would otherwise not orient themselves in the electric field. This observation may be explained by the presence of the conditioned medium factor (CMF), a protein secreted by the cells, when they begin to starve. The results of this study give new insights into understanding the mechanism that triggers the electrical sensing and transduces the external stimulus into directed cell migration. Finally, the observed increased mobility of cells over time in an electric field could offer a novel perspective towards wound healing assays.
\vspace*{1.3cm}
\end{abstract}
\vspace*{1cm}
\maketitle

Electrotaxis, also known as galvanotaxis, is the directed migration of biological cells in a DC electric field. Since it was first described over a century ago \cite{Dineur, Verwon}, the eletrotactic behavior of various cell types, including cancer cells, neurons, fibroblast, keratinocytes, leukocytes, endothelial and corneal epithelial has been reported \cite{Djamgoz2697, Fraser5381, Searson, Cooper, Sheridan, Allen, ZhaoM, Kolega, Rapp1988, Lin2465}. Electrotaxis is thought to be involved in a wide range of physiological processes, such as embryogenesis, neuronal guidance, wound healing, and metastasis \cite{Nuccitelli77, Nuccitelli, Robinson, McCaig943, McCaig4267, Mycielska1631}. 
Recently, also \textit{Dictyostelium discoideum} (\textit{Dd}), the social amoeba well-known as a model for studying cell motility and chemotaxis \cite{haastert} has proven to be a suitable model for investigating electrotaxis \cite{Zhao2002, Zhao2006, Sato2007}. However, the mechanism triggering the local activation of the signal transduction cascade that leads to actin polymerization and membrane protrusion and more generally the mechanism underlying the directed cellular movement of \textit{Dd} cells in the electric field still awaits clarifications. \newline 
A study on the involvement of cAMP receptors using cAR1$^{-}$- cAR3$^{-}$ cells showed that \textit{Dd} cells, which are unable to sense cAMP, are electrotactically as efficient as wild type cells \cite{Zhao2006}. The same study showed that the cAMP binding transduction unit constituted by the $G\alpha2$ subunit and $G\gamma \beta$ complex does not play any role in the transduction of the extracellular electric signal into directional movement. Indeed, like cAR1$^{-}$- cAR3$^{-}$, $G\alpha2^{-}$ and $G\beta^{-}$ mutants also exhibit sustained electrotaxis albeit with a reduced migration speed \cite{Zhao2006}. 
Another molecular study identified the genes required for the directional switching of electrotactic migration. 
By genetically modulating both guanylyl cyclases (GCases) and the cyclic guanosine monophosphate (cGMP)-binding protein C (GbpC) in combination with the inhibition of the phosphatidylinositide 3-kinases (PI3Ks) the cells reversed their directed migration from the cathode to the anode \cite{Sato2009}. Gao et al. \cite{Gao2015} uncovered genes involved in the electrotactic response by identifying 28 strains with defective electrotaxis and 10 strains with a slightly higher directional response. They showed PiaA to be an essential mediator of electrotaxis. This gene encodes a critical component of TORC2, a kinase protein complex that transduces changes in motility by activating the kinase PKB. Furthermore, they identified several genes that encode other components of the TORC2-PKB pathway (gefA, rasC, rip3, lst8, and  pkbR1) in playing important roles in the signalling pathway for electrotaxis.
While genes and proteins that mediate electrical sensing and define the migration direction have been investigated in \textit{Dd} cells, the cellular kinematic effects caused by electric field, as well as the initial trigger mechanism of electrical sensing still present many riddles. \newline
In this study we characterize the effect of electric fields on cells in terms of migration velocity and directionality as a function of time. In addition, we introduce the concept of electrical persistence to investigate how cells invert their trajectory when the electric field is reversed. 
We also studied the response of vegetative \textit{Dd} cells and observed that the presence of conditioned medium helps them to sense the electric field and orient themselves towards the cathode. We focus our attention on the conditioned medium factor (CMF), a protein that \textit{Dd} cells secrete when they begin to starve, as a possible trigger of the cellular electrical sensing. CMF has been shown to coordinate aggregation by regulating several aspects of cAMP signal transduction such as the activation of Ca$^{2+}$ influx, adenylyl cyclase, GCases, and gene expression \cite{Yuen1251, Clarke1995}.  Besides influencing cAMP signalling, CMF also participates in regulating cell shape. Moreover, cell migration relies on pseudopod formation, and CMF appears to allow cells to create pseudopodia more frequently than cells without CMF in their surroundings \cite{Gomer2011}. CMF is therefore a reliable candidate for such a triggering task.

\section*{Materials and methods}
\begin{figure*}
	\centering
	\includegraphics[width=0.9\linewidth]{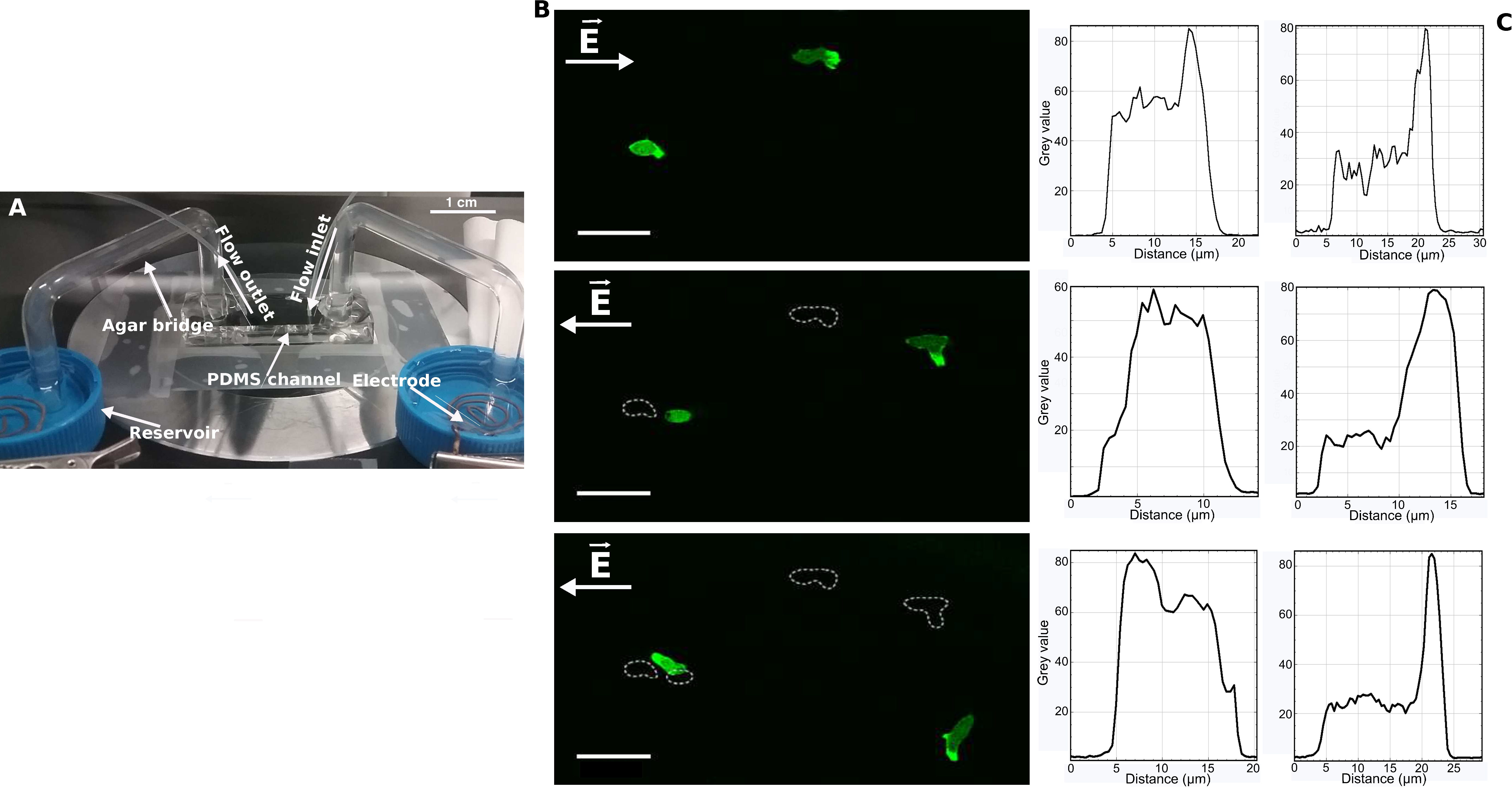}
	\caption{\textbf{Experimental set up and F-actin polymerisation}. A. Experimental set up. The electric field in the channel is generated through an indirect contact with the electrodes. The buffer reservoirs provide the electric connection between electrodes and agar bridges (for more technical details refer to Materials and Methods section). B. F-actin polymerisation through LimE-GFP. Cells with different morphology use different strategies to reverse their trajectory when the polarity of the electric field changes. The cell on the left rounds, extends a pseudopod in the new direction and forms a new leading edge whereas the cell on the right reverses its trajectory by a doing U-turn. The dashed lines show the changing position of the two cells. Scale bar: 25 $\mu$m. C. Histograms of fluorescence intensity within the cell representing the localization of F-actin. The left and right graphs correspond to the cells on the left and the right of the picture, respectively. The change in position of F-actin from the front to the back in the cell on the left when the electric field is reversed is clearly visible, while in the U-turn cell the F-actin remains at the front.}
	\label{fig:setup}
\end{figure*}

\paragraph{\textbf{Cell preparation.}}
All cell lines were derived from the axenically growing strain \textit{Dictyostelium discoideum} AX2. Wild type, LimE-GFP, ACA$^{-}$, and Amib$^{-}$ were cultivated in HL5 medium (Formedium) at 22 \textdegree C on polystyrene Petri dishes or shaken in suspension at 150 rpm. For preparation of  experiments, cells were starved in shaking phosphate buffer (PB, 2 g $\mathrm{KH}_2$$\mathrm{PO}_4$ and 0.36 g $\mathrm{Na}_2$$\mathrm{HPO}_4$$\cdot{\mathrm{2H}_2\mathrm{O}}$ per 1 L, pH 6) for different durations according to the corresponding experiment. For assays with developed cells, they were starved for 1h, 5h, 6h, 8h at a density of $\sim$ { 2}$\times$$10^6$ cells/mL. The shaking culture was pulsed with 50 nM cAMP (Sigma) every 6 min over the course of the starvation time when the experiment required it. After the corresponding starvation time with or without cAMP pulses, the cells were harvested and washed in PB. An aliquot of the cell suspension was injected into the chamber for the electrotactic assay, and the cells were allowed to spread on the glass substrate for 15 min at 22 \textdegree C. During this time a PB flow of 30 $\mu$l/h was applied to the cells in order to wash away the cAMP produced by the cells. It was switched to 50 $\mu$l/h during the experiments. 
For experiments with vegetative AX2 cells the cells cultivated in HL5 medium were detached from the Petri Dish bottom, washed twice with PB and put directly into the experimental chamber without any additional starvation time. For the experiment with conditioned medium AX2 cells were shaken for 1 h without the addition of cAMP. They were centrifuged, the buffer was harvested and centrifuged twice again to eliminate possible cells. The vegetative cells after having been detached from the Petri Dish were washed and afterwards resuspended in the conditioned buffer.
\paragraph{\textbf{Microfluidic device.}}
Figure~\ref{fig:setup}-A illustrates the design of the custom-made microfluidic device used during the experiments. Standard soft lithography was used to produce microfluidic channels 1.5 mm wide, 100 $\mu$m high, and 30 mm long. 
A master mold was fabricated transferring via photholitography a pattern to a layer of photoresist (SU-8, Micro Resist Technology), spin coated on Si wafer. To obtain the microfluidic device, polydimethylsiloxane (PDMS, 10:1 mixture with curing agent, Sylgard 184, Dow Corning Europe SA) was poured onto the wafer and cured for 45 min at 75 \textdegree C. A PDMS block containing the channels was cut out, inlets and outlets for the PB washing flow were punched through the PDMS by using a syringe tip and two holes with a diameter of 6 mm were punched at the two ends of the channels in order to insert the agar bridges. A glass coverslip (24x60 mm, $\# 1.5$, Menzel Gl{\"a}ser) was sealed to the PDMS block after a 20 sec treatment in air plasma (PDC 002, Harrick Plasma) to close the channels.
\verb   

\paragraph{\textbf{Electric connection.}}The channel was connected to the power supply through 2\% agar salt bridges 13 cm long. They were prepared in custom-made glass tubes with an internal diameter of 3 mm (Glasgeraetebau Ochs, Germany) filled with PB supplemented with 2\% (w/v) agar. One side of the agar bridge was inserted into the 6 mm diameter hole at the end of the channel, the other side was placed into  buffer reservoirs filled with PB. Ag/AgCl electrodes were immersed into the reservoirs in order to close the circuit channel-power supply. The Ag/AgCl electrodes (12 cm) were prepared by immersing two silver wires (99,9\%, 1mm in diameter, Windaus Laborthechnik, Germany) in household bleach for one hour. A programmable switch device (Siemens) was set to reverse the polarity of the electric field every 30 min within a few milliseconds. The resistance of the channel was 840 K$\Omega$ and the flowing current 25 $\mu$A. 
Upon the application of direct current voltage of 10 V/cm, we measured that an electric field of 7 V/cm was applied to the cells. This voltage drop was due to the agar bridges that we used to connect the channel with the power supply generator in order to avoid the harmful effects of the electric field on the cells, i.e. ions generated by electrolysis, changes in pH value, air bubble formation.
\paragraph{\textbf{Microscopy and cell mobility analysis.}}
The cells in the microfluidic channel were observed with an inverted microscope (Olympus IX-71) in bright field with a DeltaVision imaging system (GE Healthcare), while the migration of the cells were recorded with a CCD camera (CoolSnap HQ2, Photometrics). Data acquisition started 20 sec after the application of electric field. Cell images were acquired every 20 s for 2 hours. In the experiments with cells developed for 5-8 h cell centroids were determined manually, and the trajectories of the cell centroids were traced using MTrackJ, an ImageJ plugin. The trajectory velocity was calculated by dividing the total path length of cell migration by the time interval. The cellular velocity was presented as the mean value of the velocities of the cells recorded in each 30-minute time interval. Directionality of a cell with respect to the electric field, representative of the efficiency of the cell to migrate toward the cathode, was defined as $\cos\theta$, where $\theta$ is the angle between the vector connecting the starting and ending point of the cell trajectory and the field line of the electric field. It was calculated every minute. For the visualization of the F-actin localization, the cell contours were automatically detected using the method described in \cite{Amselem}. The center of mass of cells CM  and the intensity-weighted center of mass CM$_{W}$ were computed considering the intensity of the pixels representing the actin localization. Every 20 seconds we considered that cells were moving toward the direction of the electric field when the vector CM-CM$_{W}$ was pointing towards the cathode. When that vector was pointing in the opposite direction, we considered the cells moving anti-parallel to the electric field (see Fig. S1). The bins in Fig \ref{fig:change}-B represent the number of cells moving parallel or antiparallel to the electric field at each time point. 
Also in the experiments with vegetative and briefly starved cells, cell contours were automatically detected using the method described in \cite{Amselem}. The cell centroid was then computed and with a time interval of 20 sec between subsequent frames its position was tracked using a custom-made MATLAB program.
\section*{Results and discussion}
In this study we present results on the movement of \textit{Dd} cells in DC electric fields. We focus our analysis on the effects of electric fields on the cell kinematics and on the cellular tendency to maintain the direction of motion imposed by the electric field. We show how these effects are reflected in the actin polarization process. Lastly we present the electrotaxis of vegetative and briefly starved cells and suggest a possible mechanism regulating the electric sensing independently from the cAMP-induced development.
\subsection*{Electrotaxis of developed starved cells}
We investigated the response to the electric field of AX2 starving wild-type cells developed for 5 hours under cAMP pulsing prepared with the same development procedure used for the study of chemotaxis (see Material and Method section). The cells were seeded into a microfluidic channel where the electric field was applied for two hours. The geometry of the channel guaranteed a uniform electric field with parallel field lines along the channel (Fig. \ref{fig:setup}-A). Under the influence of an electric field of 7 V/cm cells polarized and exhibited migration towards the cathode.  We chose not to apply voltages as high as reported by other studies (up to 20 V/cm) \cite{Zhao2002, Zhao2006, Gao2011,  Gao2015} in order to reduce the Joule heating and the associated temperature increase towards non-physiological conditions (see Material and Methods). We observed a high cellular death rate for voltages above 13 V/cm; as a result, we restricted our investigation to 7 V/cm. After exposing the cells to 7 V/cm for two hours we observed no change in cellular viability and behaviour. This we tested by removing the electric field and the flow in the microfluidic device after the experimental time and verified that subsequently the cells aggregated and followed the natural life cycle.\newline
In order to remove the cooperative effects of cAMP signalling of the wild-type \textit{Dd} cells, a flow of phosphate buffer (PB) with a flow speed of 100 $\mu$m/s \cite{B801331D} parallel or antiparallel to the electric field lines was applied. The successful removal of any extracellular signalling molecule was confirmed by the absence of cell aggregates during the experiments, which would occur for \textit{Dd} cells developing normally. The flow speed was adjusted such that the mechanical shear had no effect on the directed motility of the cells.
In our case we calculated the shear stress to be 6 mPa, a value well below the critical shear stress for mechanotactical response (0.8 Pa) \cite{Decave2002,Decave2003}. We also verified the effect of electro-osmotic flow on the cell motility. We analysed the movement of beads with a diameter of approx. 1 $\mu$m in the flow induced by electro-osmosis. The velocity of the beads was found to be around 8.8 $\mu$m/s, i.e., much smaller compared to the velocity of the external applied flow. Therefore the effect of the electro-osmotic flow on the cell migration can be neglected.
Altogether, this shows  that the directed migration elicited by the electric field in our experiments was not influenced by external factors such as the applied flow, electro-osmotic flow, or chemical gradients.\newline
We also found no evidence that extracellular Ca$^{2+}$ is necessary for electrotaxis. The experiments presented here were conducted with Ca$^{2+}$-free phosphate buffer. Moreover, the flow in the microfluidic device washes away any compound released by the cells. This evidence contradicts the results by  Shanley et al. \cite{Shanley4741}, where it was found that electrotaxis was not possible in the absence of extracellular Ca$^{2+}$. \newline
When the electric field polarity was reversed, the cells turned around and migrated towards the new cathode (See Movie S1), confirming the observations reported by \cite{Sato2007, Zhao2002}. 
In response to inverting the field polarity the cells reversed their trajectory by using two different strategies, depending on their initial morphology: In most cases polarised cells with a single pseudopod or an accentuated pseudopod achieved reorientation by making a U-turn while maintaining their morphological polarity. Cells with a less polarised initial morphology reoriented by forming extensions of the pseudopod in the direction of the new cathode, i.e. reversing the front and the rear. With LimE-GFP as an in-vivo marker for F-actin we visualised the distribution of actin during this process. Fig \ref{fig:setup}-B (see also Movie S2) shows an example of cells with the two different cellular morphologies: the U-turn cell shows strong localisation of F-actin at the leading pseudopod, while it is less pronounced in the other cell. In response to the polarity change of the field, the U-turn cell retains its F-actin localisation, while in less polarized cells F-actin is redistributed from back to front, forming  a new leading edge. This behaviour was also observed during chemotaxis \cite{Haastert2004} and demonstrates that the electric field triggers migration by activating the molecular signalling pathway that transduces an external stimulus into actin polymerisation. Therefore, an intracellular electric polarization due to passive electrostatic effects of the electric field can be excluded.
\subsection*{Cellular velocity, directionality and temporal electric persistence}
\begin{figure}[h!]
	\centering
	\includegraphics[width=.9\linewidth]{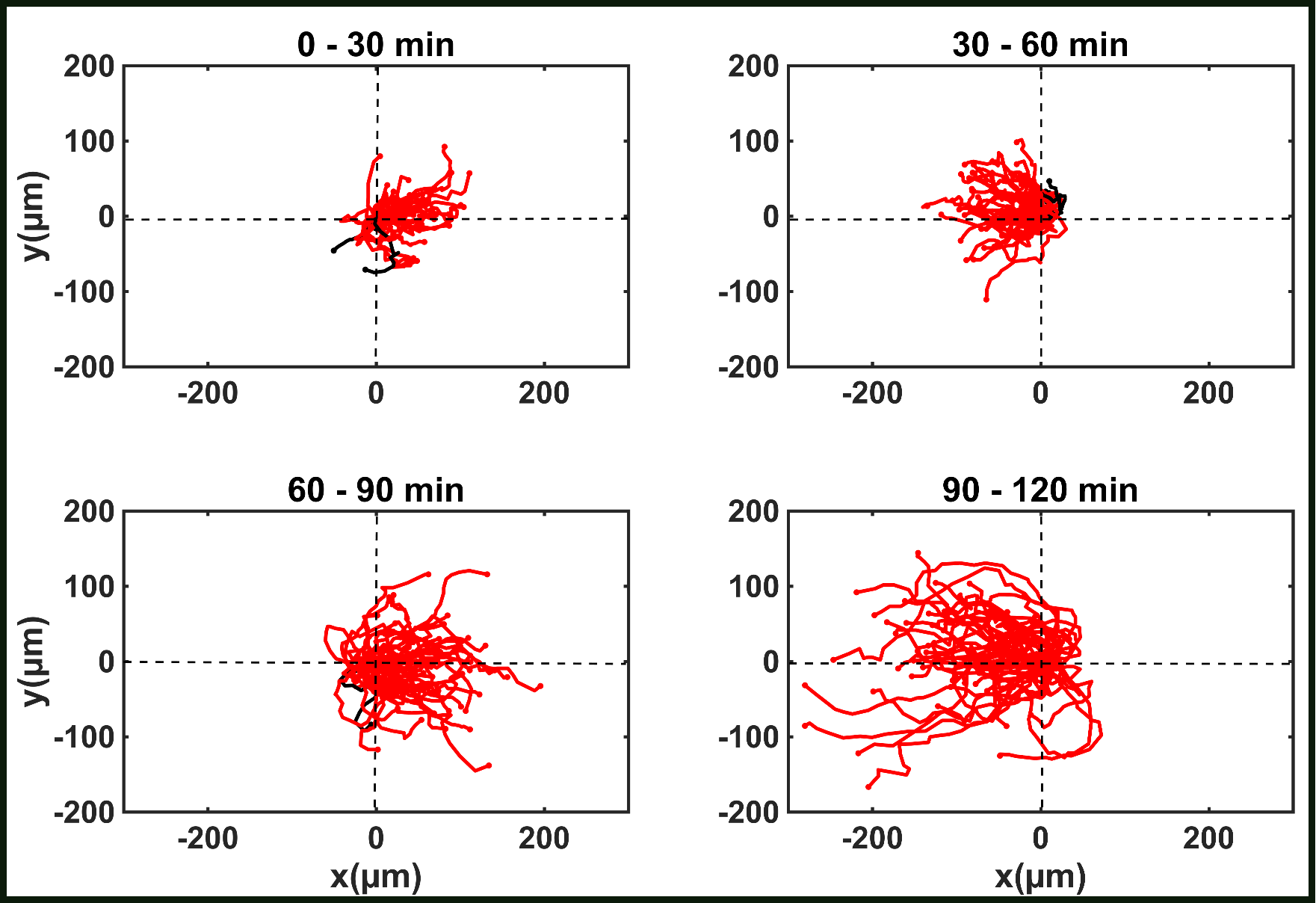}
	\caption{\textbf{Cell tracking diagrams of fully developed AX2 cells}. The length of the migration paths increases significantly over time. In each experiment at least 45 cells were analysed. The diagrams refer to cells that were starved for 5 hours. They originated from at least three different experiments. 
	}
	\label{fig:5hstarv}
\end{figure}

\begin{figure}[h!]
	\centering
	\includegraphics[width=0.9\linewidth]{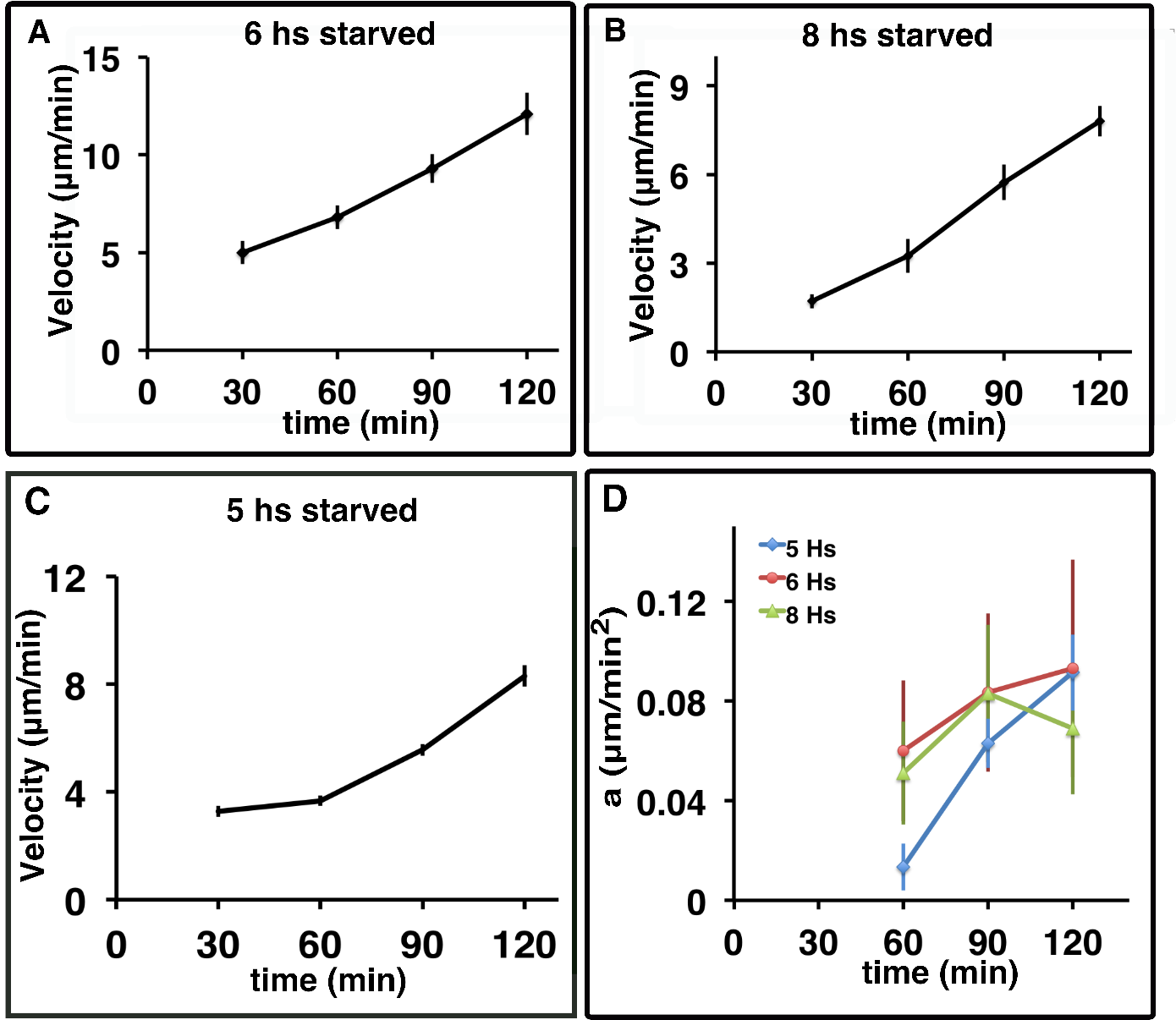}
	\caption{\textbf{Cell acceleration}. A. Migration velocity of Ax2 cells starved for 6 hours. B. Migration velocity of Ax2 cells starved for 8 hours. C. Migration velocity of Ax2 cells starved for 5 hours. D. Cell acceleration over time for cells starved and pulsed 5, 6 or 8 hours. All data are represented as $mean \pm{ s.e.m}$}
	\label{fig:accel}
\end{figure}

\begin{figure}[h!]
	\centering
	\includegraphics[width=0.9\linewidth]{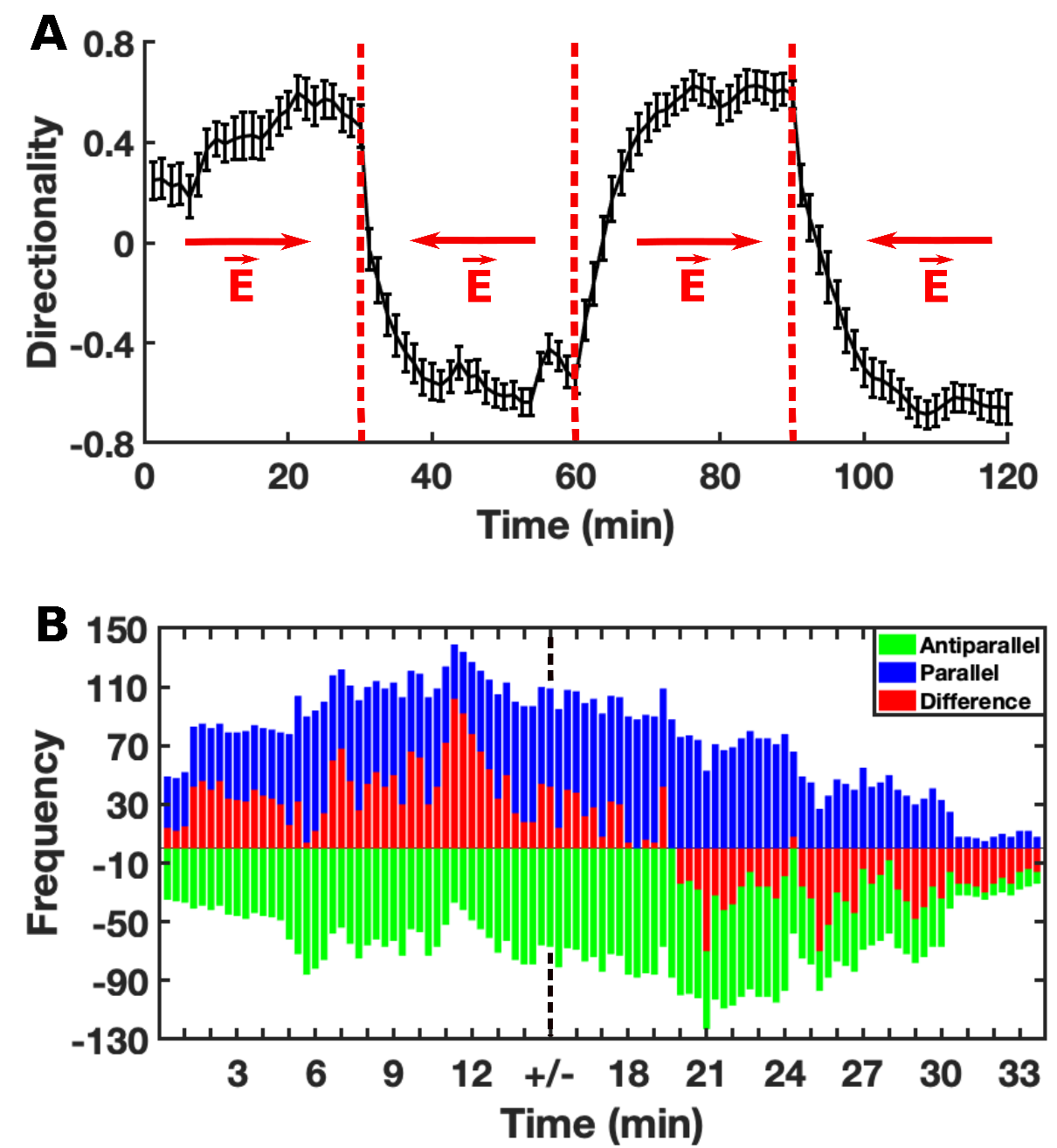}
	\caption{\textbf{Directionality and F-actin localization upon electric field reversal} A. Directionality of \textit{Dd} cells in an electric field. The polarity of the field is reversed every 30 minute as indicated by the red arrows. The graph is obtained from at least three different experiments. The data are displayed as $mean \pm{ s.e.m}$. B. F-actin localization within the cell during cellular migration and its dynamics when the polarity of electric field changes. At each point in time, the number of cells with F-actin localized towards the cathode (blue), towards the anode (green) and their difference (red) are displayed. The black dotted line indicated the time point of polarity reversal.}
	\label{fig:change}
\end{figure}

\begin{figure}[h!]
	\centering
	\includegraphics[width=0.9\linewidth]{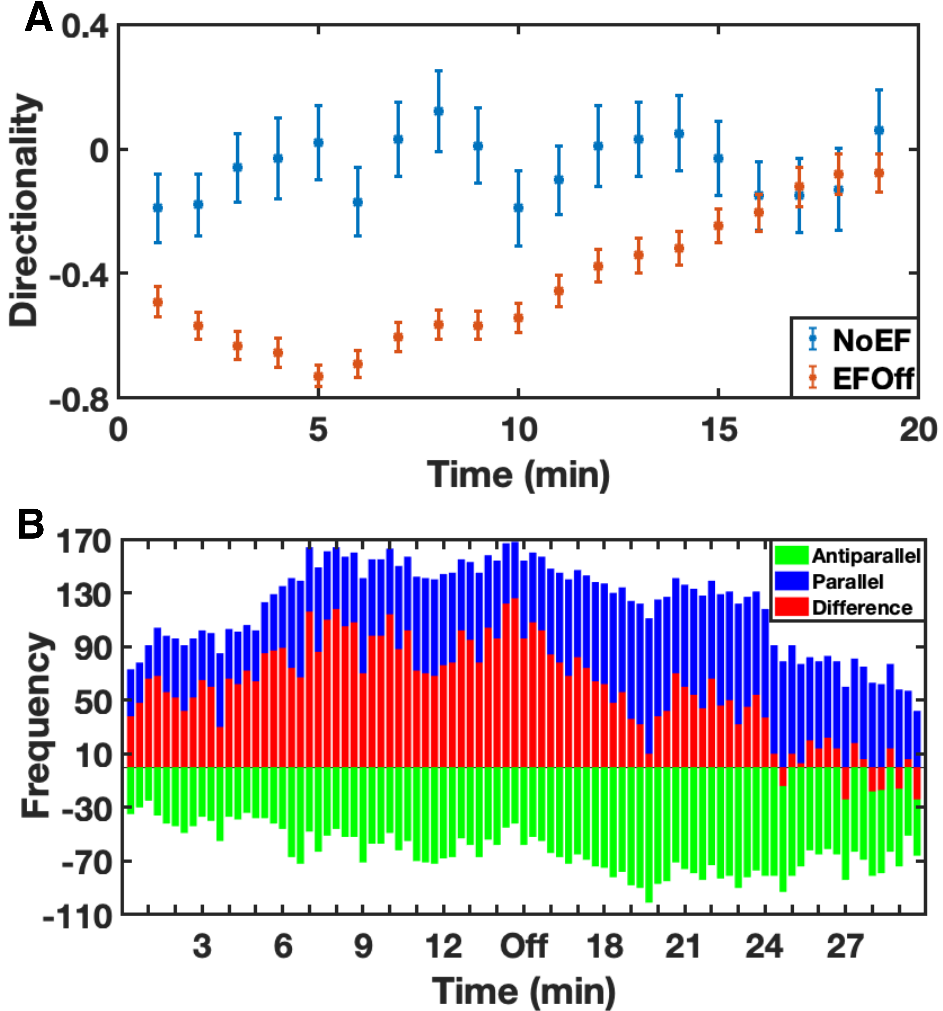}
	\caption{\textbf{Electrical persistance without electric field} A. Directionality of \textit{Dd} cells that were never exposed to electric field and move randomly (blue) and cells where the electric field was switched off after 90 min (red). The electric field was removed at t=10 min. It is clear that the directionality decreases during the time after the field removal. The data are represented as $mean \pm{ s.e.m}$. B. F-actin localization inside the cell during the cellular migration and its dynamics when the electric field is switched off. }
	\label{fig:EFoff}
\end{figure}

When the cells were exposed to the electric field for two hours, with the field reversing every 30 minutes, we observed that the migratory velocity  increased continuously during the observation period. The increasing length of the cellular trajectory over time is clearly visible in Fig \ref{fig:5hstarv}, and the corresponding cell velocity increases from 3.27 $\pm$  0.2 $ \mu$m/min in the first 30 minutes to 8.3 $\pm$ 0.4 $ \mu$m/min after 2 hours (Fig~\ref{fig:accel}-C ). 
To investigate whether this observation depends on the stage of development induced by the cAMP pulsing procedure, we tested the response of cells that were starved and pulsed for 5, 6 or 8 hours. The cell speed for the case of 6 and 8 hours is shown in Fig. \ref{fig:accel}-A,B. The acceleration of the cells caused by the electric field was also observed in these cases and calculated as $a$ = $\Delta v$/$\Delta t$, with $\Delta v$ the difference in velocities between time intervals and $\Delta t$ the time interval of 30 minutes. Figure \ref{fig:accel}-D shows that the three cell populations reacted to the electric field with an increase of the initial velocity, regardless of their developmental stage. 
We speculate that the cAMP pathway involved in the activation of the aggregation adenylyl cyclase (ACA) and associated production of cAMP could play a significant role. In fact, we observed that two mutant strains unable to produce cAMP and aggregate, namely ACA$^{-}$ and Amib$^{-}$, did exhibit electrotaxis, but did not show any increase in velocity  over time (See S2). Nevertheless at this stage this assumption is a speculation and only a detailed molecular study similar to \cite{Gao2015} can elucidate the true reason for the speed up. 
\newline
Interestingly, the directionality of the cells did not change significantly (Fig \ref{fig:change}-A) reaching a plateau value ${Dir_{max}}$, which ranged from 0.54 in the first 30 min to 0.65 in the last interval.
The transition phase between two plateaus provides information about the reaction of cells to the reversal of the polarity of electric field. 
By fitting this cells response to the exponential function ${Dir_{max}} - c e^{\frac{t}{\tau}}$ we calculated that it is characterized by a time constant $\tau_{30-60}$= 2.94 min in the first electric field reversal, followed by $\tau_{60-90}$= 5 min and $\tau_{90-120}$= 5.64 min in the second and third one, respectively (where the subscript of $\tau$ refers to the corresponding time interval in minutes). It appears clear that the longer the cells are under the influence of the electric field, the longer it takes for them to reverse their trajectory when the polarity of the electric field is  reversed. Thus, the adaptation to their environment prevents the cells from reacting immediately to any change in the electric field and rearranging the migratory machinery.
To characterize this behavior, we defined the temporal electrical persistence or electrical memory, i.e. the time required by cells to reverse their trajectory in an inverted electric field. In this way, we could gain insights into the cellular sensing process and its transduction into the molecular pathway. For this purpose we analyzed the localization of the F-actin network by using LimE-cells and we could observe how the adaptation behavior was reflected by  F-actin polymerization inside the cell. The cells were kept in the electric field for 90 minutes. Afterwards, the polarity was reversed and the cells were observed for 20 minutes (Fig.~\ref{fig:change}-B). We evaluated the occurrence of cells in which F-actin was localized towards the cathode and of cells in which it was localized in the opposite direction. The difference between these two populations represents the net population of cells polarized in the direction defined by the electric field (see section Material and Method for analysis details). These results show that the cells need 4.3 min $ \pm$ 20 sec to shift the F-actin in the new direction of the electric field after the polarity change. \newline
To better characterized the sensing mechanism, we analysed the temporal electrical persistence of the cells when the electric field was removed and tested how long it took for the cells to``forget'' the environmental stimulus and move randomly. Cells migrating without any electrical stimulus showed no preferred direction (Fig.~\ref{fig:EFoff}-A), and the average directionality of their random movement was 0 $ \pm$ 0.2.  
We measured the temporal electrical persistence of cells without electrical influence after they were kept in the electric field for 90 minutes (Fig.~\ref{fig:EFoff}-A). We saw that the cells without electrical influence reduced their directionality towards the electric field and showed random movement after a period of 6 minutes. By visualizing the F-actin we could observe that the time needed to completely lose polarization and not to show preferred localization of F-actin was 9.2 min $ \pm$ 20 sec (Fig.~\ref{fig:EFoff}-B). Also in this case we could observe that the directed movement decreased with time when the electric stimulation was switched off.
\subsection*{Electrotaxis of initially vegetative cells}
\begin{figure}[h!]
	\centering
	\includegraphics[width=.8\linewidth]{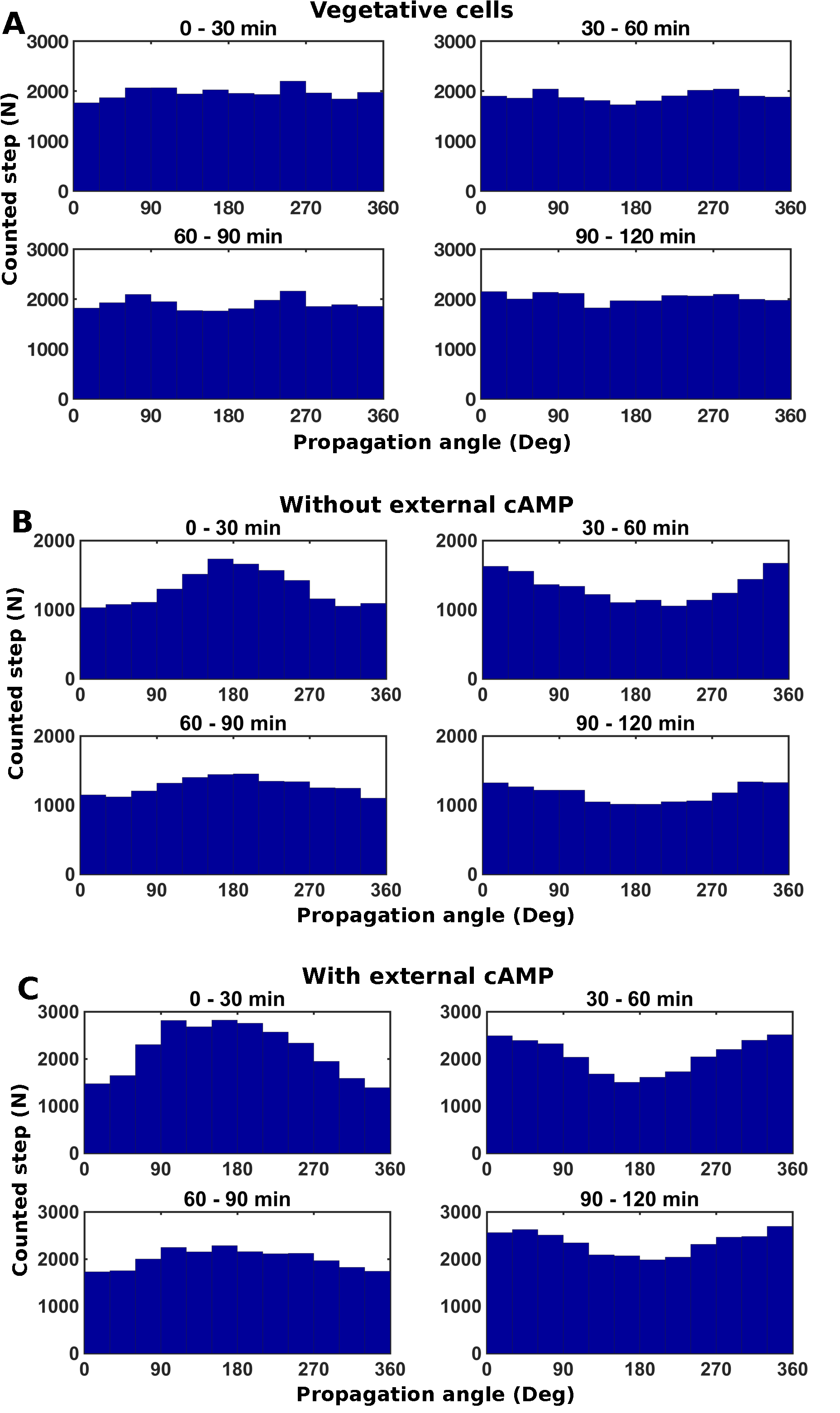}
	\caption{\textbf{Histograms of the propagation angle distribution}. Histograms of the propagation angle distribution for vegetative cells (A), briefly starved cells in a shaking culture without (B) and with (C) external cAMP pulses. Cell motion is biased towards the cathode with $\theta$ = 180$^\circ$ in the first 30 min and $\theta$ = 0$^\circ$ or 360$^\circ$ when the polarisation of the electric field has been reversed. The histograms represents the distribution of the propagation angle $\theta \in [0^\circ, 360^\circ] $ of each time step with respect to the electric field lines pointing towards the cathode. Every histogram resulted by the analysis of 100 - 300 cells from three different experiments.}
	\label{fig:30min}
\end{figure}

In this study we were also interested to understand the mechanism involved in the initiation of the 
electrical sensing in cells. For this purpose we analysed the behaviour of vegetative and briefly starved \textit{Dd} cells under the influence of electric fields. It allowed us to study the response of cells before they entered the development phase and went though the program of gene-expression changes induced by cAMP pulses.\newline
Vegetative wild type AX2 cells cultivated in HL5 medium were washed and resuspended in PB, then seeded immediately into the microfluidic device. We analysed their behaviour under the influence of electric fields over a period of up to 7 hours, during which the polarity of the electric field was reversed every 30 minutes. No directed movement was observed; rather, the cells migrated randomly during the whole observation time. Figure \ref{fig:30min}-A shows the behaviour of vegetative cells in the electric field for 2 hours. After 7 hours deprived of nutrients in PB the cells did not show the typical characteristics of starved cells in an electric field such as morphological changes and migration towards the cathode. As a control experiment, we tested the effect of the electric field on cell physiology and repeated the experiment by keeping the cells under the flow for 5 hours without electric field and then applying the electric field for 2 hours. Again, the cells showed no preferred direction of movement. In order to check the involvement of calcium (Ca$^{2+}$) in this behaviour we repeated the experiments by substituting the Ca$^{2+}$-free PB for a buffer containing 1mM CaCl. No remarkable difference was observed in the cells response to the electric field. We concluded that under these experimental conditions the cells were not able to enter the development stage normally induced by starvation. Thus, entering the developmental phase and activating genes related to the starvation program are necessary steps for enabling cells to sense and respond to the electric field. 
\subsection*{Electrotaxis of briefly starved cells} 
The influence of the initial starvation on the behaviour of cells in electric fields was investigated by experiments with cells starved for 1 hour both with and without additional cAMP pulses in shaking culture (see Materials and Methods). We observed that cells starved for 1 hour with and without additional cAMP pulses reacted similarly by migrating or realigning their cell body towards the cathode when the electric field was applied for 30 minutes. They inverted their orientation towards the new cathode when the polarity of the electric field was reversed (Fig. \ref{fig:30min}-B-C, see also S3). Interestingly, after one hour the directed movement decreased with time, independent of the change in the polarization of the electric field. As in the case of vegetative cells, we repeated these experiments using a buffer containing 1mM CaCl as washing out flow. Again, no difference in the attenuation effect after 1 hour was observed. \newline
These results show that the induction of development by nutrient deprivation leads to an increased electrotactic capacity, but not the presence of exogenous cAMP. This is not surprising since Yuen at al. have shown that the ability of cells to transduce external cAMP occurs 2 hours after starvation \cite{Yuen1251}.
\subsection*{Conditioned Medium restores the cellular sensing for external electric stimuli}
\begin{figure}[h!]
		\centering
		\includegraphics[width=.9\linewidth]{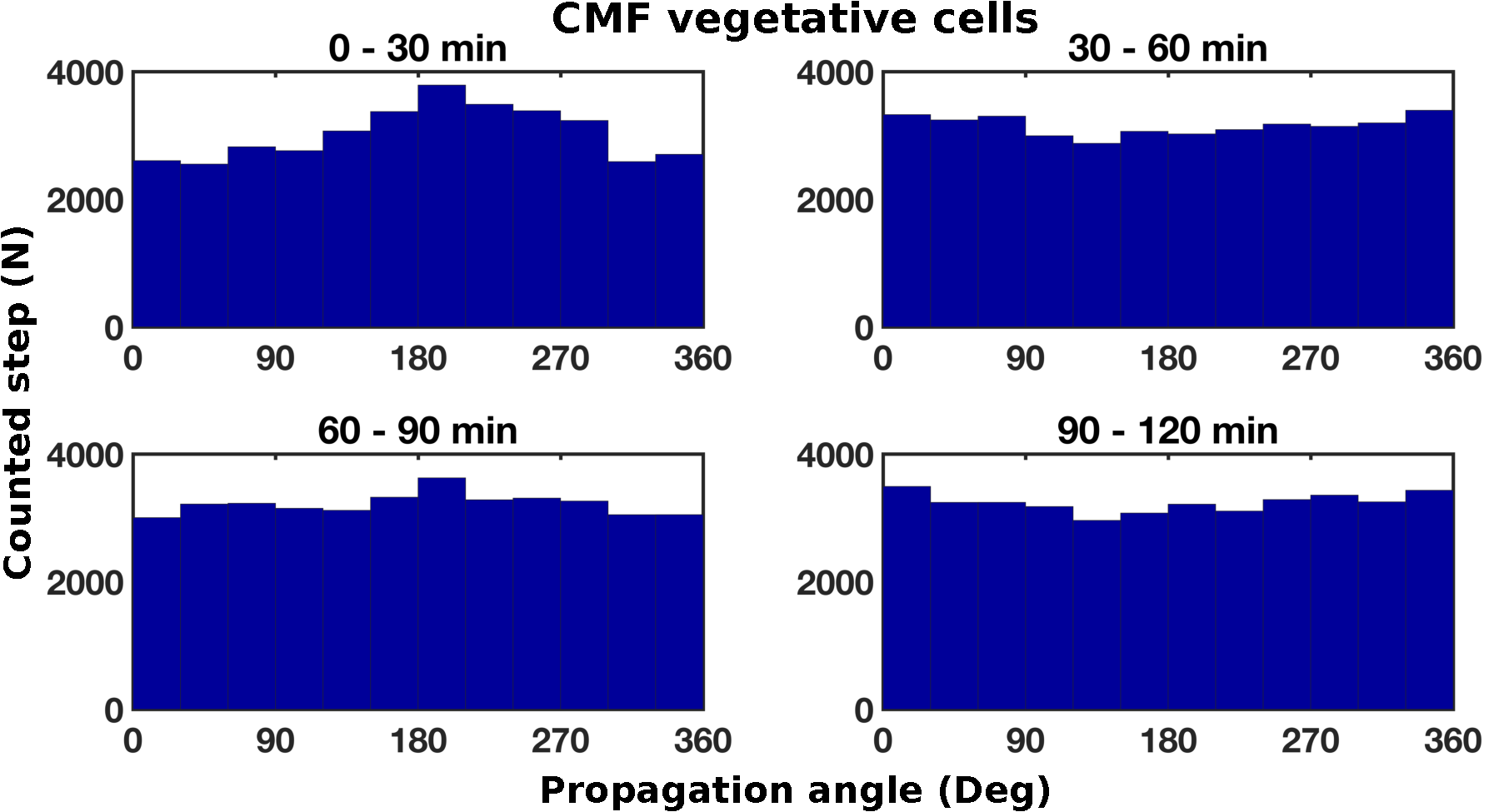}
		\caption{\textbf{Histograms of the propagation angle distribution for conditioned vegetative cells.} Every histogram resulted by the analysis of 300 - 350 cells from three different experiments}
		\label{fig:30minCM}
	\end{figure}

	So far we have observed a difference in electrical sensing between vegetative and 1-hour-starved cells. We verified that the presence of external cAMP is not essential for triggering the electrical sensing of cells. We therefore assume that the electrical sensing is triggered by a signaling molecule that the cells immediately release at the onset of starvation. In order to test this hypothesis, we studied the reaction of vegetative cells to electric fields when they were resuspended in conditioned medium. We starved wild-type cells in a shaking culture for 1 h, centrifuged them, removed the conditioned buffer and used this conditioned medium to resuspend vegetative cells. These conditioned vegetative cells were immediately seeded into the microfluidic channel and conditioned medium was also used as washing flow instead of PB during the entire observation period of 2 hours.
	Interestingly, under these conditions the conditioned vegetative cells showed an electrotactic motion and reorient their cellular body towards the cathode. Fig. \ref{fig:30minCM} clearly shows the reaction of the cells to the electric field. The involvement of conditioned medium in the electrical sensing and its triggering effect is clearly visible when comparing the cellular response of the initially vegetative cells (Fig. \ref{fig:30min}-A) and the conditioned vegetative cells (Fig. \ref{fig:30minCM}). By comparing the reactions to the electric field of the conditioned vegetative cells (Fig. \ref{fig:30minCM}) and the briefly starved cells (Fig. \ref{fig:30min}-B-C), we can see that conditioned medium rescues the sensing of cells for electric fields, but the cells react for a shorter time. 
Although there are dozens of autocrine agents in the conditioned medium of cells starved for only one hour, we speculate that the decisive factor triggering the cellular electrical sensing is the conditioned medium factor (CMF) \cite{Yuen1251}, a protein that the cells release at the onset of starvation.
It is secreted throughout development by \textit{Dd} cells, but not by vegetative cells \cite{Gomer269}. Various studies suggest that CMF is essential for cell-cell communication and for the coordination of cell aggregation. It acts during the earliest stages of starvation and is used by the cells to enter the developmental phase and induce the expression of selected genes depending on the spatial density of cells \cite{Clarke1995, Yuen1375}. \newline
While conditioned vegetative cells immediately showed electrotaxis, vegetative cells were not able to sense the electric field when seeded in a channel with a PB flow up to 7 hours. We attribute this behavior to the flow that washes away not only cAMP but also the CMF secreted by the cells. Under these conditions they cannot enter the developmental phase. 
According to our hypothesis,  CMF receptors might be responsible for the signal transduction of the external electric field and thus for the activation of the signaling pathways that lead the cells to polarization and directed migration. It is known that the gene encoding CMF receptors is expressed in vegetative cells and CMF receptors accumulate on the cell membrane when starvation sets in \cite{Jain25031994}. In addition, CMF binds to CMF receptors and regulates PldB \cite{Gomer2011}, a phospholipase D homologue known to regulate actin localization \cite{zouwail} and pseudopod formation \cite{Gomer2011}, which are essential for cell migration.
This is consistent with the fact that in our microfluidic experiments vegetative cells did not show any electrotaxis, while briefly starved cells and conditioned vegetative cells were electrotactic. The temporal decrease in electrotactic directionality in all three cases may be attributable to the lack of increase in CMF.
\section*{Conclusion}
In this study we present results on the electrotactic behavior of \textit{Dd} cells at different developmental stages. We showed that fully developed wild-type \textit{Dd} cells respond to the electric field by increasing their migratory velocity over time. However, the mutant strains ACA$^{-}$ and Amib$^{-}$ migrate at a constant velocity over the observation period. At this stage it is unclear whether the enzyme ACA, which is known to be essential for oscillatory cAMP signals, is involved in the determination of the electrotactic migratory velocity. We introduced the temporal electrical persistence as the time delay of cells to respond to the variation of the external electric field. We quantified the time interval in which the cells rearrange their migratory machinery either to reverse their migratory trajectory or to ``forget" the electric stimulus after switching off of the field. This provides insight into the mechanism by which cells transduce the sensing of the electric stimulus into a kinematic response.
By analysing vegetative cells that are not capable of directed movement, we show that the electric sensing can be rescued by using conditioned medium. We believe that among the autocrine molecules that the cells release at the onset of starvation, CMF is the prime candidate for triggering the electrical sensing. Only when cells have been exposed to CMF (self-produced or supplied with conditioned medium) do they respond to the electric field with directed migration. This assumption requires an analysis using the protein CMF instead of the conditioned medium. Until now, the protein could not be made available to test our assumption. We encourage researchers to follow this line of inquiry.
\\
\paragraph{\textbf{Acknowledgments}}
  E.B. and I.G. acknowledge support from the MaxSynBio Consortium which is jointly funded by the Federal Ministry of Education and Research of Germany and the Max Planck Society. We acknowledge Albert Bae, Narain Karedla, and Christian Westendorf for fruitful discussion and suggestions; Maren M\"uller, Katharina Gunkel, Achim Hillebrandt, and Riccardo Guido for technical support.

\end{document}